 \documentclass{scrartcl} 

\usepackage[english]{babel}
\usepackage[utf8]{inputenc}
\usepackage{lmodern}

\usepackage{bbm}
\usepackage{amsmath,amssymb,amsthm}
\usepackage{amssymb}
\usepackage{booktabs}
\usepackage{graphicx}
\usepackage{url}
\usepackage[round]{natbib}
\usepackage[paper=a4paper,left=30mm,right=30mm,top=30mm,bottom=30mm]{geometry}

\newcommand{\intd}{\textnormal{d}}
\newcommand{\ind}{\mathbbm{1}} 
\newcommand{\cN}{\mathcal{N}}
\newcommand{\R}{\mathbb{R}}

\title{Comparison of nonhomogeneous regression models for probabilistic wind speed forecasting}

\author{Sebastian Lerch\footnote{Institute of Applied Mathematics, Heidelberg University, Im Neuenheimer Feld 294, 69120 Heidelberg, Germany}\hphantom{ } and Thordis L. Thorarinsdottir\footnote{Norwegian Computing Center, P.O. Box 114, Blindern, 0314 Oslo, Norway}}

\date{\today}

\begin{document}

\maketitle

\begin{abstract}
In weather forecasting, nonhomogeneous regression is used to statistically postprocess forecast ensembles in order to obtain calibrated predictive distributions.  For wind speed forecasts, the regression model is given by a truncated normal distribution where location and spread are derived from the ensemble.  This paper proposes two alternative approaches which utilize the generalized extreme value (GEV) distribution.  A direct alternative to the truncated normal regression is to apply a predictive distribution from the GEV family, while a regime switching approach based on the median of the forecast ensemble incorporates both distributions.  In a case study on daily maximum wind speed over Germany with the forecast ensemble from the European Centre for Medium-Range Weather Forecasts, all three approaches provide calibrated and sharp predictive distributions with the regime switching approach showing the highest skill in the upper tail. 
\end{abstract}

\section{Introduction}

Reliable forecasts of wind speed are a necessity in a diverse number of applications such as agriculture, most modern means of transportation and wind energy production.  Wind power, as a renewable and emissions free alternative to fossil fuels, has been growing rapidly over the last decade.  In Europe, the wind power's share of total installed power capacity amounted to about 11.4\% at the end of 2012 and it has increased five-fold since 2000 \citep{EWEA2012}.  For wind energy production, accurate forecasts of wind speed at different lead times are required to regulate electricity markets, to schedule maintenance and, more generally, to improve the competitiveness of wind power compared to sources of electricity which allow for dispatchable generation \citep{GenHer2007, PinEt2007, LeiEt2009}.  In many of these applications and for weather warnings, high wind speeds are of particular importance.

The focus of this paper are daily forecasts with medium-range lead times of 1-3 days.  In this setting, forecasts are usually based on outputs from numerical weather prediction (NWP) models which use physical descriptions of the atmosphere and oceans to propagate the state of the atmosphere forward in time based on the current weather conditions.  Moreover, to account for uncertainties in the knowledge of the initial state of the atmosphere and the numerical model, the NWP models are often run several times with different initial conditions and/or numerical representations of the atmosphere resulting in an ensemble of forecasts \citep{GneRaf2005, LeuPal2008}.   The development of ensemble prediction systems plays a key role in the transition from deterministic to probabilistic forecasting and has become an established part of weather and climate prediction \citep{Pal2002}. 

Probabilistic forecasts are essential in many applications in that they allow for quantification of the associated prediction uncertainty. Further, optimal decision-making requires probabilistic forecasts, particularly for rapidly fluctuating resources such as wind energy where the optimal point forecast depends on permanently changing market features \citep{PinEt2007,ThoGne2010}.  See \citet{Gne2011} for a detailed discussion of optimal deterministic forecasts based on probabilistic forecasts.   While ensemble systems are valuable in this context, they are finite and do not provide full predictive distributions. Also, they tend to be underdispersive and subject to a systematic bias, and thus require some form of statistical postprocessing \citep{HamCol1997,GneRaf2005,GneEt2007}.  
  
Statistical postprocessing methods for ensembles of wind speed forecasts include the ensemble Bayesian model averaging (BMA) method of \citet{SloEt2010} and the nonhomogeneous regression (NR) or ensemble model output statistics (EMOS) approach developed by \citet{ThoGne2010}.  BMA predictions are given by weighted mixtures of parametric densities or kernels each of which depends on a single ensemble member, with the mixture weights determined based on the performance of the ensemble members in the training period.  For wind speed, \citet{SloEt2010} apply a mixture of gamma distributions, see also \citet{CouEt2013}.  The NR method of \citet{ThoGne2010}, on the other hand, applies a single normal distribution truncated at zero, where the location parameter is an affine function of the ensemble members and the scale parameter is an affine function of their variance.  In a comparison study, the two methods show very similar predictive performance \citep{ThoGne2010}.  The NR method has been extended to wind gusts \citep{ThoJoh2012} and a BMA approach for wind direction is proposed in \citet{BaoEt2010}.  \citet{Pin2012}, \citet{SchEt2012} and \citet{SloEt2013} study statistical postprocessing of bivariate wind vector ensembles. 

Hourly average wind speeds are usually modeled using lognormal, gamma, Rayleigh or Weibull densities, with the Weibull model showing the best performance in many case studies, see for example \citet{GarEt1998} and \citet{Cel2004}.  Here, we consider forecasts of daily maximum wind speed and the predictive distributions are conditioned on the ensemble forecast, the situation for which the postprocessing approaches of \citet{SloEt2010} and \citet{ThoGne2010} were developed.  Since daily maximum wind speeds are block maxima, results from extreme value theory imply that the generalized extreme value (GEV) distribution provides a suitable model \citep{Col2001}. GEV distributions have especially received attention in modeling maxima of wind and gust speed observations over long return periods, typically 50 years, see \citet{PalEt1999} and references therein.  \citet{FriTho2012} apply a GEV model for probabilistic predictions of daily peak wind speed.  

\begin{figure}
  \centering
  \includegraphics[width=0.45\textwidth]{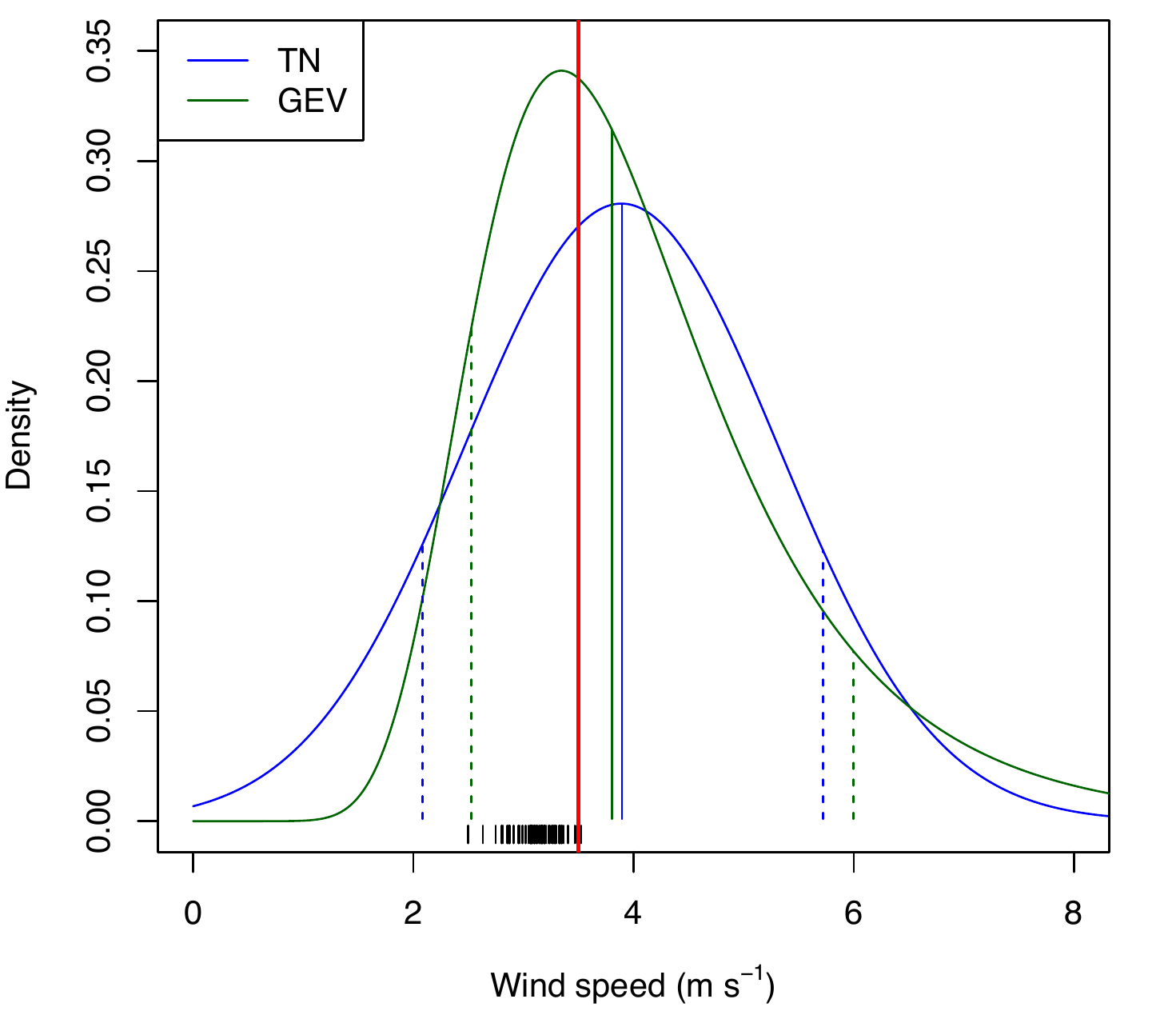}
  \caption[Predictive distributions at Frankfurt Airport on 19 March 2011.]{One day ahead forecasts for daily maximum wind speed at Frankfurt Airport valid on 19 March 2011. The ECMWF ensemble forecasts are indicated in black, the observation in red.  The solid blue and green lines indicate the median of the truncated normal (TN) and the GEV predictive distribution, respectively.  The dashed lines indicate the corresponding central 80\% prediction intervals.}
  \label{fig:example}
\end{figure}
  
We propose to combine the NR ensemble postprocessing framework originally proposed by \citet{GneEt2005EMOS} and later extended by \citet{ThoGne2010} with results from extreme value theory.  That is, we apply a predictive distribution from the GEV family, where the location and scale parameters depend on the ensemble member forecasts.  To illustrate the difference between the two NR approaches, Figure \ref{fig:example} shows the predictive distributions for Frankfurt Airport on 19 March 2011.  Both the truncated normal (TN) and the GEV model correct the negative bias and the underdispersion of the ensemble, while the GEV density is less symmetric and exhibits a heavy right tail.  We further investigate a regime switching approach which issues a TN predictive density when the ensemble median forecast takes a low value and a GEV predictive density for high values of the ensemble median. 
  
The remainder of the paper is organized as follows. In Section 2, the ensemble forecasts and the observational data are introduced.  In Section 3, we review the NR technique and describe our extensions using GEV distributions and a regime switching combination model.  Section 4 summarizes the probabilistic scores used for estimating the model parameters and evaluating the competing forecasting procedures.  In particular, we discuss how appropriately weighted proper scoring rules recently proposed in the economic literature \citep{GneRan2011,DikEt2011} can be used to assess the predictive performance for high wind speeds.  In Section 5, we report the results of a case study on daily maximum wind speed over Germany for lead times of 1-3 days with the ensemble issued by the European Centre for Medium-Range Forecasts from May 2010 to April 2011.  We close with a discussion in Section 6.

\section{Data}

 \begin{figure}
 \centering
   \includegraphics[width=0.35\textwidth]{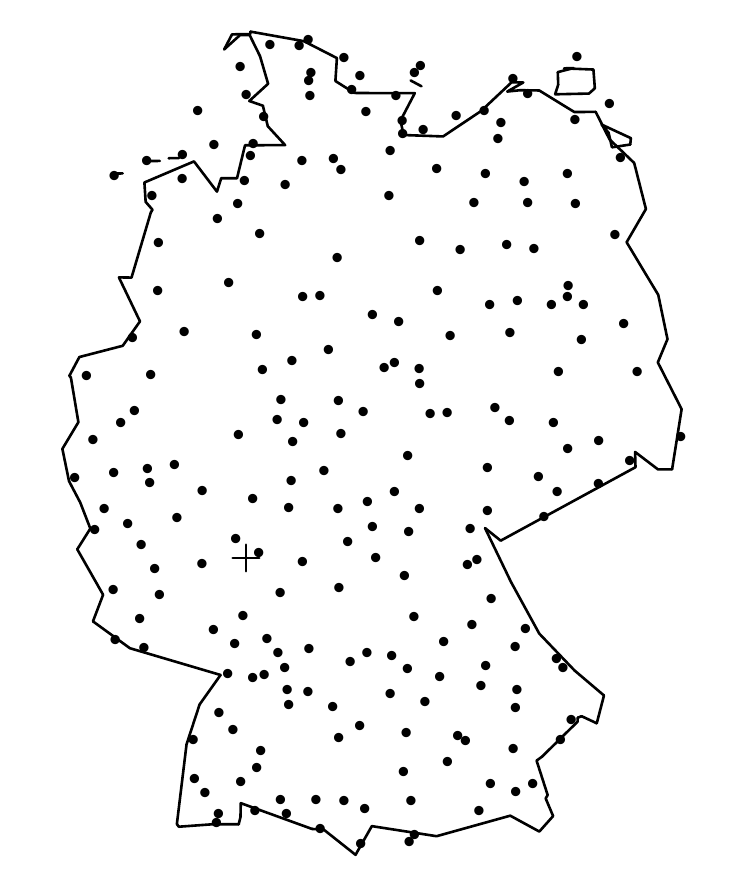}
   \caption[Map of Germany showing the locations of the 228 synoptic observation stations.]{Map of Germany showing the locations of the 228 synoptic observation stations used in this study.  The station at Frankfurt Airport is indicated with $+$.}
   \label{fig:map}
\end{figure}
 
We consider 50 ensemble member forecasts of near-surface (10-meter) wind speed obtained from the global ensemble prediction system of the European Centre for Medium-Range Weather Forecasts (ECMWF).  Ensemble forecasts for lead times up to 10 days ahead are issued twice a day at 00 UTC and 12 UTC, with a horizontal resolution of about 33 km and a temporal resolution of 3-6 hours.  To account for uncertainties in the initial conditions and the numerical model, the ensemble members are generated from random perturbations in initial conditions and stochastic physics parametrization \citep{MolEt1996, LeuPal2008, PinHag2012}.  The ensemble members are thus statistically indistinguishable and can be treated as exchangeable \citep{FraEt2010}.  We restrict attention to the ECMWF ensemble run initialized at 00 UTC and lead times of 1-3 days.  To obtain predictions of daily maximum wind speed, we take the daily maximum of each ensemble member at each grip point location.  For instance, one day ahead forecasts are given 
by the maximum over lead times of $3, 6, \ldots, 24$ hours.  

The forecasts are verified over a set of 228 synoptic observation stations over Germany, see Figure \ref{fig:map}.  The observations are hourly observations of 10-minute average wind speed which is measured over the 10 minutes before the hour.  To obtain daily maximum wind speed, we take the maximum over the 24 hours corresponding to the time frame of the ensemble forecast.  In the following, the terms wind speed and daily maximum wind speed are used synonymously.  Ensemble forecasts at individual stations are obtained by bilinear interpolation of the gridded model output.  The results presented below are based on a verification period from 1 May 2010 to 30 April 2011, consisting of 83\,220 individual forecast cases.  Additionally, we use data from 1 February 2010 to 30 April 2011 to obtain training periods of equal lengths for all days in the verification period and for model selection purposes.

  \section{Nonhomogeneous regression prediction models}

The nonhomogeneous regression (NR) methodology was originally developed for sea level pressure and surface temperature under a normal predictive distribution \citep{GneEt2005EMOS}, see also \cite{HagEt2008} and \cite{KanEt2009} for further applications.  \cite{ThoGne2010} extend the framework to wind speed using a normal distribution truncated in zero while \cite{ThoJoh2012} apply the same set-up to predict gust speeds based on NWP forecasts of wind speed and gust factors.  A bivariate normal model for wind vectors is discussed in \cite{SchEt2012}.  An NR framework for quantitative precipitation has recently been proposed by \cite{Sch2013} using a GEV model for the precipitation accumulations. 

\subsection{Truncated normal model}\label{sec: TN}
  
Let $Y$ denote wind speed and $X_1,\dots,X_k$ the corresponding ensemble member forecasts. The predictive distribution for $Y$ is given by a truncated normal (TN) distribution with a cutoff at 0,
\begin{equation}\label{eq:TN}
  Y|X_1,\dots,X_k \sim \cN_{[0,\infty)}(\mu,\sigma^2),
\end{equation}
where the location parameter $\mu = a + b_1X_1 + \dots + b_k X_k$ is an affine function of the ensemble forecasts and the variance $\sigma^2 = c + d S^2$ is an affine function of the ensemble variance $S^2 = \frac{1}{k} \sum_{i=1}^k (X_i - \bar{X})^2$ with $\bar{X} = \sum_{i=1}^k X_i$.  As the ECMWF ensemble members are exchangeable, we assume that $b_1 = \dots = b_k$, or $\mu = a + b \bar{X}$ \citep{FraEt2010}.  The distribution function of the TN distribution is given by
\[
F(z) = \Phi\left(\frac{\mu}{\sigma}\right)^{-1} \Phi\left(\frac{z-\mu}{\sigma}\right)
\]
for $z>0$ and 0 otherwise, where $\Phi$ denotes the cumulative distribution function of the standard normal distribution.

\subsection{Generalized extreme value model}

As an alternative to the TN model in (\ref{eq:TN}), we consider a model based on extreme value theory.  The cumulative distribution function of the GEV distribution with location parameter $\mu$, scale parameter $\sigma$ and shape parameter $\xi$ is given by
\begin{equation}\label{eq:GEV}
  G(z) = \begin{cases}
    \exp \left\{ - \left[ 1+\xi\left( \frac{z-\mu}{\sigma} \right) \right]^{-1/\xi} \right \} & \xi \neq 0 \\
    \exp\left\{-\exp \left[ - \left( \frac{z-\mu}{\sigma} \right) \right] \right\}		  & \xi = 0.
  \end{cases}
  \end{equation}
This distribution  is defined on the set $\{z\in\R : 1+\xi(z-\mu)/\sigma > 0\}$, where the parameters satisfy $\mu,\xi \in\R$ and  $\sigma > 0$.  For $\xi > 0$, $G$ is of Fr\'echet type with a heavy right tail and it holds that $z \in [\mu - \sigma / \xi, \infty)$.  We obtain the Fr\'echet type in approximately $99.5\%$ of our forecast cases.  We estimate the parameters of the model in (\ref{eq:GEV}) without any constraints on the parameter values. It is thus possible to obtain non-zero probabilities of negative wind speed. However, we find that this rarely happens in practice.  The probability of negative wind speed is larger than $1\%$ in about $0.1\%$ of the forecast cases and it never exceeds $5\%$. 

To link the parameters of the predictive GEV distribution to the ensemble, we apply the Bayesian covariate selection algorithm described in \cite{FriTho2012}.  We assume a constant shape parameter $\xi$ while the location $\mu$ and the scale $\sigma$ may depend on the ensemble mean and variance.  Based on an analysis of the out-of-sample data from 1 February to 30 April 2010, we let the parameters depend on the ensemble mean only and set $\mu = \mu_0 + \mu_1 \bar{X}$,  $\sigma = \sigma_0 + \sigma_1 \bar{X}$.  While the average posterior inclusion probability for the ensemble mean is very high for both location and scale, the average posterior inclusion probability for the ensemble variance amounts to less than $0.1\%$ for both parameters.  The results were obtained for 100\,000 iterations of the Metropolis within Gibbs algorithm with a burn-in period of 20\,000 iterations.

\subsection{Regime switching combination model}
  
The third model we consider is a regime switching method which combines the TN approach in (\ref{eq:TN}) and the GEV approach in (\ref{eq:GEV}).  Conditional on the median of the ensemble predictions,
\[
X^{med} = \textnormal{median}(X_{1},\dots,X_{k}),
\]
we either issue a TN or a GEV predictive distribution. That is, for a model threshold $\theta\in\R_+$, we define the predictive distribution by
\begin{equation}\label{eq:RS}
H =  \begin{cases}
  \mathcal{N}_{[0,\infty)}\left(\mu^\mathcal{N},\sigma^{2\,\mathcal{N}}\right), & \textnormal{if } X^{med} < \theta \\
    G\left(\mu^{G},\sigma^{G},\xi^{G}\right), & \textnormal{if } X^{med} \geq \theta.
\end{cases}   
\end{equation}

Here, the parameters of the TN and GEV model depend on the ensemble forecast as described above.  However, we train the TN model only on training data for which it holds that $X^{med} < \theta$.  Similarly, the parameters of the GEV distribution are learned from data where $X^{med} \geq \theta$.  The model threshold $\theta$ is selected by comparing predictive performance over a range of possible thresholds based on the out-of-sample data from 1 February to 30 April 2010.  Generally, thresholds between 7 and 8 m s$^{-1}$ prove optimal which approximately corresponds to the 75th and 85th percentile of the median ensemble predictions over the verification period.  These results are discussed in detail below.  Under this model, the probability of negative wind speed is less than $1.4 \times 10^{-5}$ for all forecast cases.  

\section{Parameter estimation and prediction verification}

The aim of the prediction is to ``maximize the sharpness of the predictive distribution subject to calibration'' \citep{GneEt2007}. Calibration is a joint property of the predictive distribution and the associated observation. It essentially requires that the observation is indistinguishable from a random draw from the predictive distribution.  Sharpness refers to the concentration of the predictive distribution and is a property of the forecasts only.  

\cite{And1996} and \cite{HamCol1997} propose verification rank (VR) histograms as a graphical tool to assess the calibration of ensemble predictions.  VR histograms show the distribution of the ranks of the observations when pooled within the ordered ensemble predictions.  For a calibrated ensemble, the observations and the ensemble predictions should be exchangeable, resulting in a uniform VR histogram.  The continuous analog of the VR histogram is the probability integral transform (PIT) histogram \citep{Daw1984, GneEt2007}.  The PIT is the value of the predictive cumulative distribution function at the corresponding realizing observation.  Again, for calibrated forecasts, the PIT values should follow a uniform distribution.

  \subsection{Proper scoring rules}

 Scoring rules assign numerical values to forecast-observation pairs and provide summary measures of predictive performance.  Forecasting methods can be compared in this manner by averaging their scores over a test set.  If the scoring rule evaluates the full predictive distribution, it can simultaneously address calibration and sharpness.  A scoring rule is proper if the expected score is minimized when the true distribution of the observation is issued as the forecast \citep{BroSmi2007,GneRaf2007}.  Proper scores thus prevent hedging strategies. 

Popular examples of proper scoring rules are the logarithmic or ignorance score \citep{Goo1952},

\begin{equation} \label{eq:logs-def}
  \textnormal{LogS}(F,y) = - \log(f(y)),
\end{equation}

where $f$ denotes the density of $F$ and $y$ denotes the corresponding observation, and the continuous ranked probability score (CRPS) \citep{Her2000, GneRaf2007},

\begin{equation} \label{eq:crps-def}
  \textnormal{CRPS}(F,y) =  \int_{-\infty}^{\infty} (F(z) - \ind\{ y \leq z \})^2 \intd z ,
\end{equation}

where the distribution $F$ is assumed to have a finite first moment.  Again, $y$ denotes the corresponding observation.   We furthermore use the absolute error $| x-y |$ for the point forecast $x$ given by the median of the predictive distribution as a deterministic measure of accuracy. The median of the predictive distribution is the Bayes predictor under the absolute error loss function \citep{Gne2011}.  All these scoring rules are negatively oriented, that is, the smaller, the better.

  \subsection{Evaluation of forecasts for high wind speeds}  
  Despite the variety of theoretically justifiable methods to evaluate probabilistic forecasts, it is not obvious how to assess the predictive performance in the tails of the distribution, for example in the case of extreme wind speed observations. A natural approach is to select extreme events while discarding non-extreme events, and to proceed using standard evaluation procedures. However, it can be shown that restricting proper scoring rules to subsets of events results in improper scoring rules.  This approach is thus bound to discredit even the most skillful forecasters \citep{GneRan2011}.  Instead, weighted scoring rules that emphasize specific regions of interest can be constructed. 
  
  \cite{GneRan2011} propose the threshold-weighted continuous ranked probability score (twCRPS), 
\begin{equation} \label{eq:twcrps-def}
    \textnormal{twCRPS}(F,y) = \int_{-\infty}^{\infty} (F(z) - \ind\{y\leq z\})^2 w(z) \intd z,
  \end{equation}
where $w(z)$ is a non-negative weight function on the real line. For $w(z)\equiv 1$, the twCRPS reduces to the original CRPS in \eqref{eq:crps-def}.
  If the interest lies in the right tail of the distribution, we may for example set $w(z) = \ind\{z \geq r\}$.  Note that the CRPS in \eqref{eq:crps-def} represents an integral of the Brier score \citep{Bri1950} over all possible thresholds.  The twCRPS in \eqref{eq:twcrps-def} with $w(z) = \ind\{z \geq r\}$ thus allows us to simultaneously assess the exceedance probabilities for all thresholds greater or equal to $r$.    Similarly, \cite{DikEt2011} propose proper weighted versions of the logarithmic score in \eqref{eq:logs-def}.

\subsection{Optimum score estimation}

The framework of  proper scoring rules may also be applied to parameter estimation. Following the general optimum score estimation approach of \cite{GneRaf2007}, the parameters of a distribution are determined by optimizing the average value of a proper scoring rule as a function of the parameters over a training set.  Optimum score estimation based on the logarithmic score in \eqref{eq:logs-def} thus corresponds to maximum likelihood (ML) estimation. Minimum CRPS estimation, that is, optimum score estimation based on the CRPS in \eqref{eq:crps-def}, provides a robust alternative to ML estimation if closed form expressions for the CRPS of the distribution family of interest are available.
  
  Following \cite{ThoGne2010}, we estimate the parameters of the TN model using minimum CRPS estimation.  \cite{FriTho2012} derive a closed form expression of the CRPS for the GEV distribution and compare ML and minimum CRPS estimation for their analysis of peak wind speed.  For our data set, ML estimation proved to be more parsimonious and numerically stable.  There is no analytical solution of the corresponding ML minimization problem \citep{Col2001}.  However,  numerical approximations can be obtained using standard algorithms for any given dataset \citep{PreWal1980}.   For the regime switching combination model in \eqref{eq:RS}, minimum CRPS estimation is applied for the parameters of the TN distribution, and ML estimation for the parameters of the GEV distribution.   For all three methods, the parameters are estimated over a rolling training period consisting of the forecast-observation pairs of the last $m$ days.  The parameters are estimated regionally in that training data from all stations are 
pooled together.

\section{Results}\label{sec:results}

Here, we present the results for 1-3 day ahead probabilistic forecasts of daily maximum wind speed over Germany produced by the three different postprocessing methods presented in Section 3.  The verification period covers one year,  from 1 May 2010 to 30 April 2011. 

\subsection{Selection of training period and regime switching threshold}

The results presented here are based on a rolling training period of length $m = 30$ days for all methods.  We have also performed the same analysis for training periods of length $m = 20, 25, \dots, 50$ days.  In general, shorter training periods allow for a rapid adaption to changes in environmental conditions while longer training periods reduce the statistical variability in the parameter estimation \citep{GneEt2005EMOS}.  We found that the performance scores reported in Table \ref{table:perf-measures-24h} change by less than 1\% for the different values of $m$, and in accordance with the results of \cite{ThoGne2010} and \cite{ThoJoh2012}, we conclude that the methods are robust against changes in $m$. 

\begin{figure}
 \centering
   \includegraphics[width=0.45\textwidth]{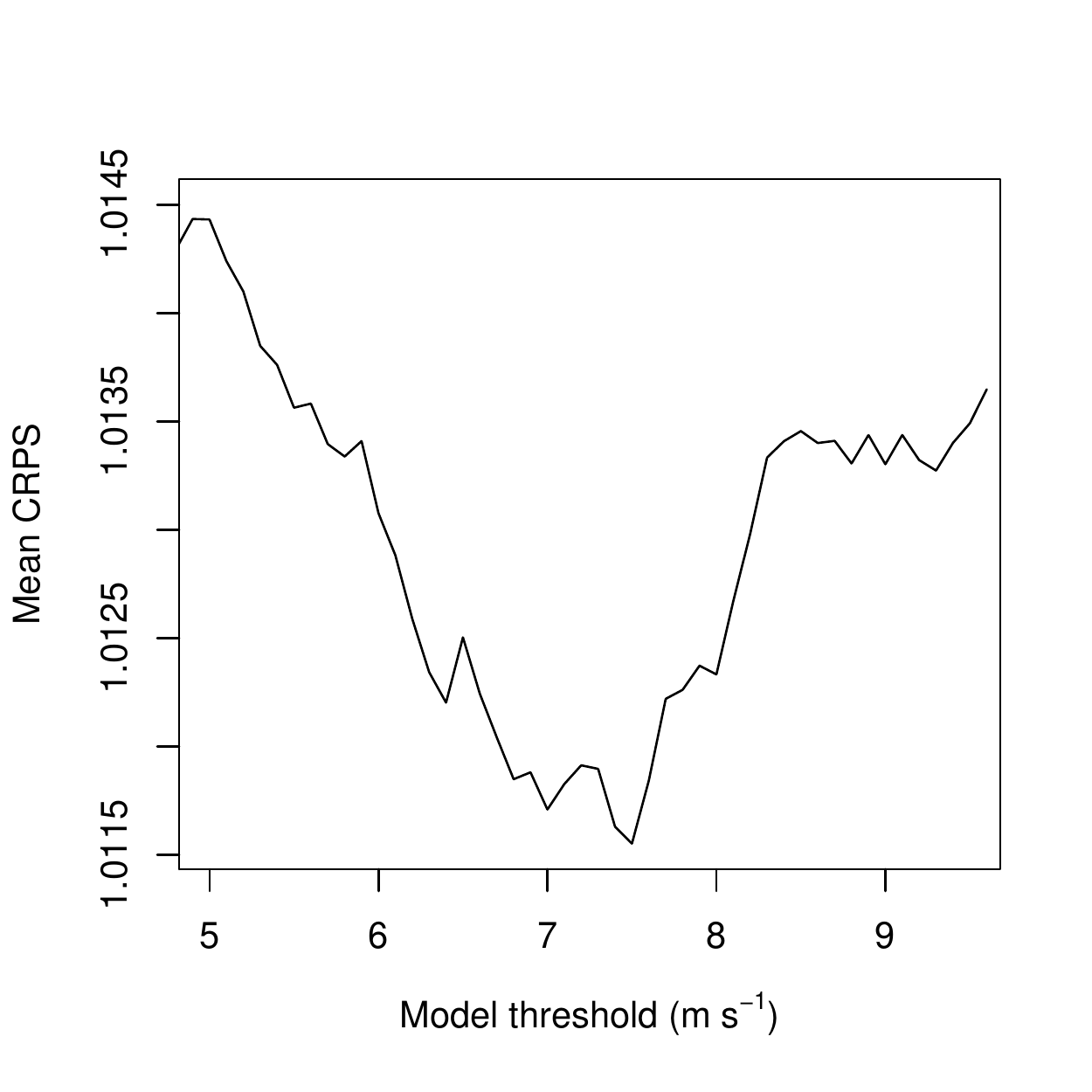}
   \caption[Determination of the model threshold for the regime switching model.]{Mean continuous ranked probability score (CRPS) for the regime switching model as a function of the model threshold $\theta$.  The results are based on a rolling training period of 30 days during the out-of-sample time period from 1 February 2010 to 30 April 2010.}
   \label{fig:determine-theta}
\end{figure}
  
The model threshold $\theta$  for the regime switching combination model in \eqref{eq:RS} is determined by computing the mean CRPS for a range of threshold values over an out-of-sample training period from 1 February 2010 to 30 April 2010.  Using a rolling training period of $m=30$ days, we obtain the optimal score for $\theta = 7.5$ m s$^{-1}$, see Figure \ref{fig:determine-theta}. A sensitivity analysis shows that the results in Table \ref{table:perf-measures-24h} are nearly constant for values of $\theta$ between 7 and 8 m s$^{-1}$ and various choices of $m$.  The threshold of $7.5$ m s$^{-1}$ corresponds approximately to the 80th percentile of the median ensemble predictions in the verification set.  Over the verification period, a GEV distribution is used in around 18\% of the forecast cases.
  
  \subsection{One day ahead predictive performance}

\begin{figure}
 \centering
   \includegraphics[width=0.4\textwidth]{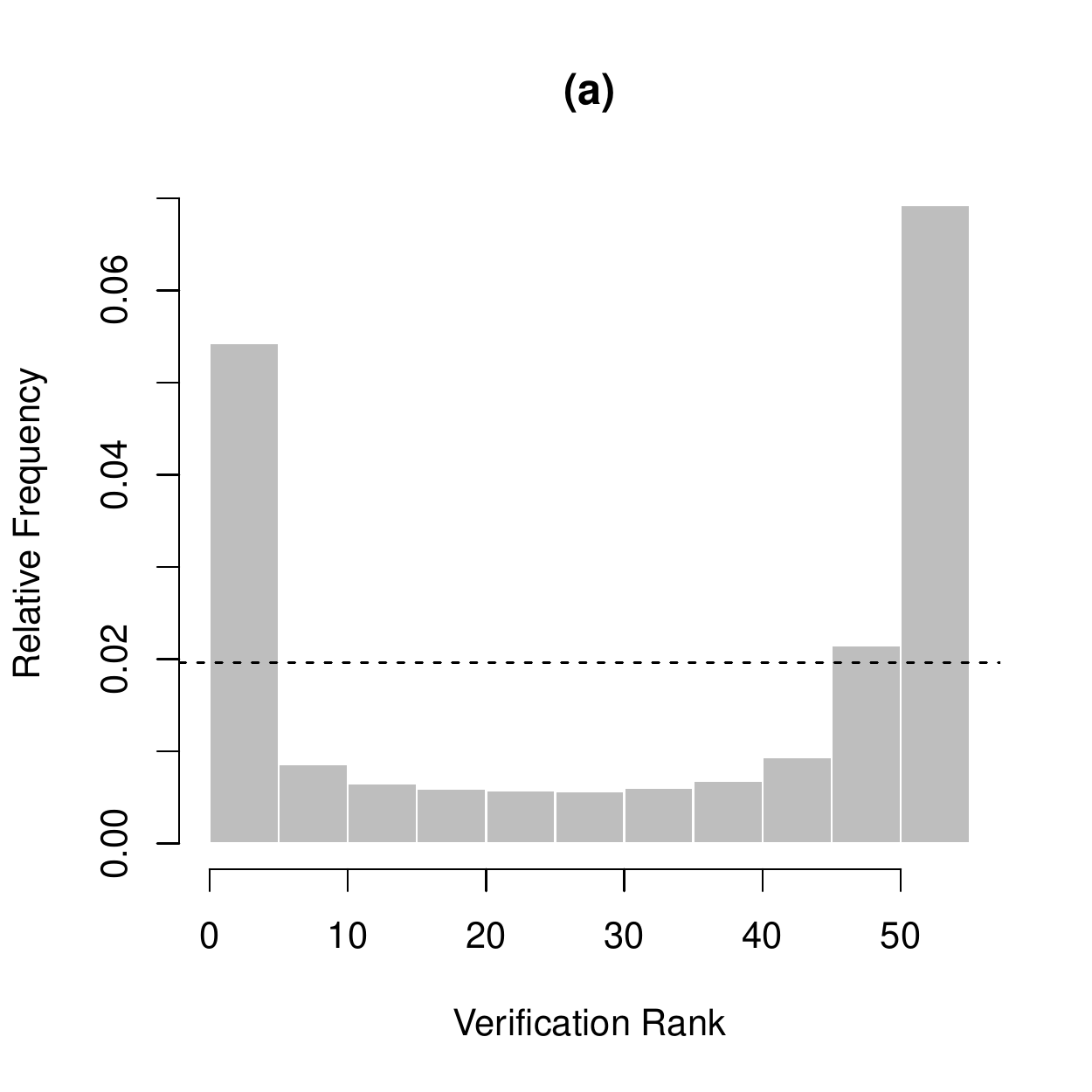}
   \includegraphics[width=0.4\textwidth]{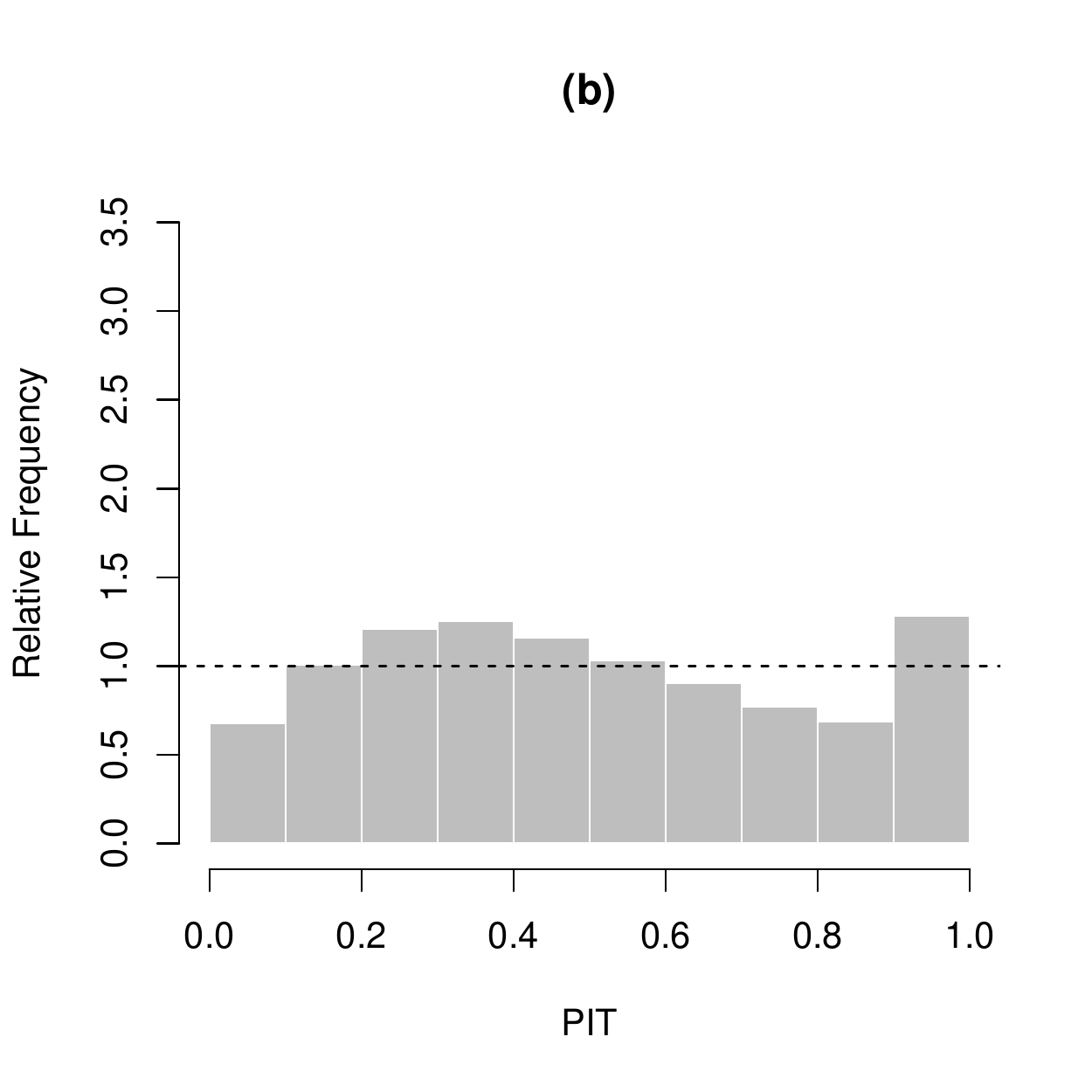}  \\
   \includegraphics[width=0.4\textwidth]{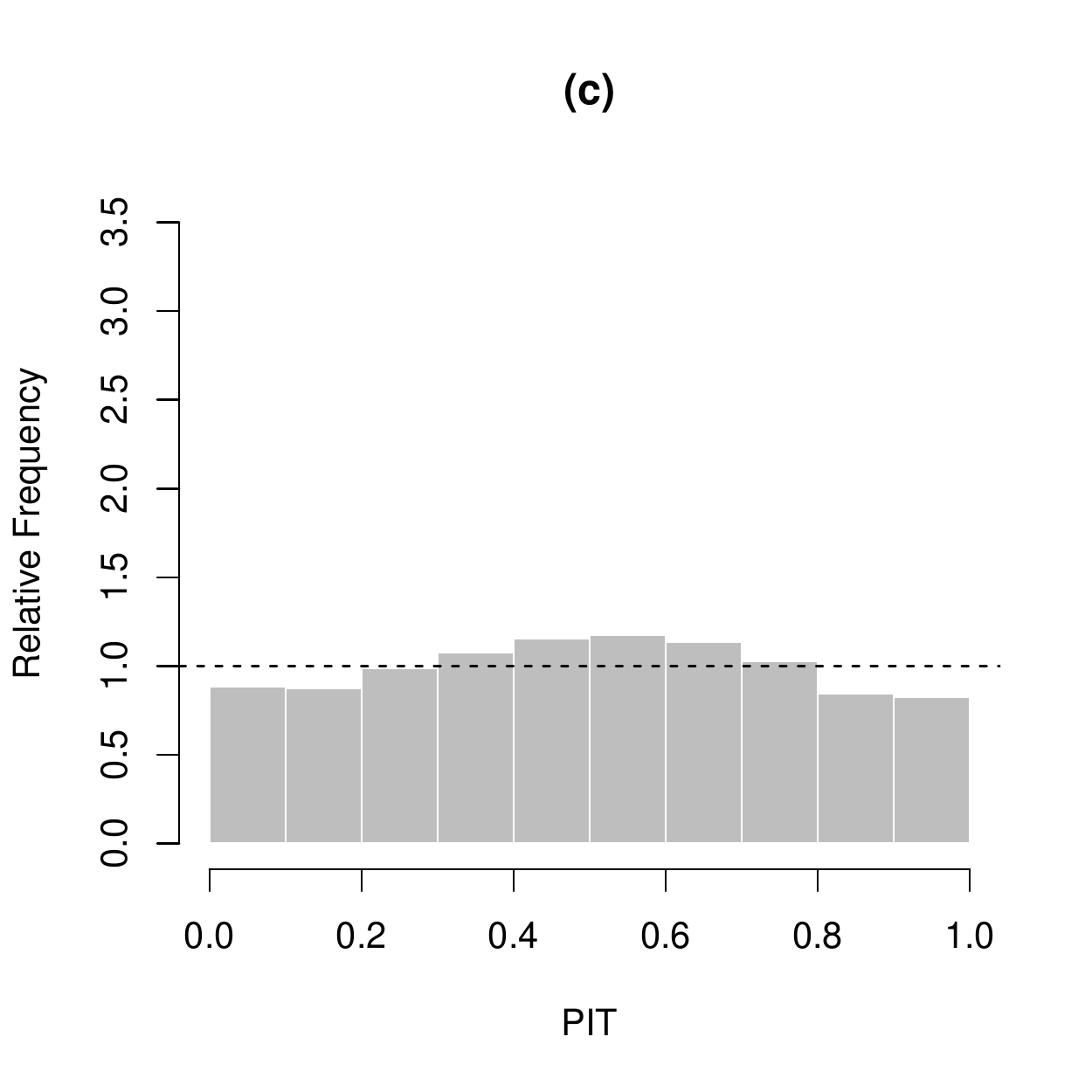}
   \includegraphics[width=0.4\textwidth]{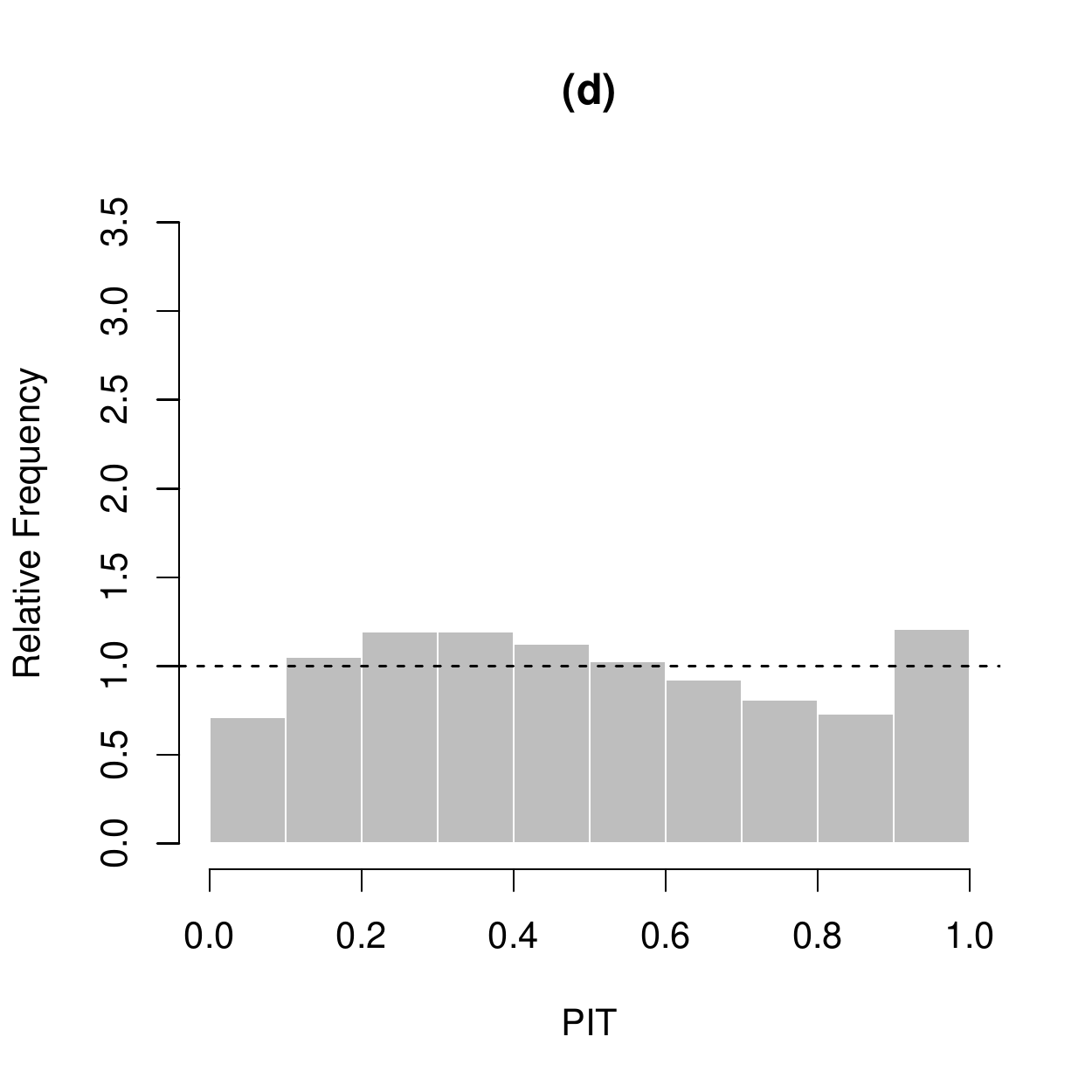}
  \caption[Calibration checks for one day ahead forecasts of wind speed.]{Calibration checks for probabilistic one day ahead forecasts of wind speed over Germany aggregated over 1 May 2010 to 30 April 2011 and the 228 stations: (a) verification rank histogram for the ECMWF ensemble forecasts; (b) PIT histogram for the TN model; (c) PIT histogram for the GEV model; (d) PIT histogram for the regime switching combination technique.}
  \label{fig:pit}
\end{figure}

  \begin{table}
  \centering
  \caption[Summary measures for one day ahead forecasts.]{Mean continuous ranked probability score (CRPS), mean absolute error (MAE), average coverage and width of 80\% prediction intervals of probabilistic one day ahead forecasts of daily maximum wind speed at 228 synoptic stations in Germany from 1 May 2010 to 30 April 2011.}
  \begin{tabular}{lcccc}
    \toprule
	    & CRPS         & MAE          & Coverage & Width \\
      Forecast & (m s$^{-1}$) & (m s$^{-1}$) & (\%)     & (m s$^{-1}$) \\
    \midrule
      Climatology & 1.54           & 2.13          & 64.4          & 6.6 \\
      Ensemble    & 1.26           & 1.44          & 26.6          & 1.0 \\
      TN        & 1.05           & 1.39          & \textbf{80.4} & 4.0  \\
      GEV         & 1.04           & 1.39          & 82.9          & 4.6  \\
      Combination &  \textbf{1.03} & \textbf{1.38} & 80.8          & 4.1  \\
    \bottomrule
    \end{tabular}
    \label{table:perf-measures-24h}
  \end{table}
  
  We compare the three ensemble postprocessing methods discussed above to the raw, unprocessed ECMWF ensemble and a climatological reference forecast. For each day, the climatological reference forecast is obtained from the observed wind speeds in the 30 day training period used for the parameter estimation of the postprocessing methods.  Verification rank and PIT histograms for the ensemble and the three postprocessing methods are shown in Figure \ref{fig:pit}. The ECMWF forecasts are underdispersive, with too many observations falling outside the ensemble range.  All postprocessing methods significantly improve the calibration of the ensemble.  While the GEV forecasts are slightly overdispersive and show smaller deviations from uniformity than the TN forecasts, the PIT histogram of the combination model resembles the PIT histogram of the TN technique, with minor improvements for large PIT values. 

Table \ref{table:perf-measures-24h} shows the mean CRPS, the MAE and average coverage and width of 80\% prediction intervals for the competing forecasts.  For calibrated forecasts, the average coverage of 80\% prediction intervals should be close to 80\% and narrower average prediction intervals indicate sharper forecasts. For discrete distributions such as the ECMWF ensemble and the climatology forecast, the CRPS can be calculated explicitly, see e.g. \cite{BerEt2008}.  The CRPS for the TN model and the GEV model is calculated as described in \cite{ThoGne2010} and \cite{FriTho2012}, respectively.  The ECMWF ensemble predictions outperform the climatological reference forecast and provide sharp prediction intervals at the cost of being uncalibrated. All postprocessing methods outperform the ensemble predictions, with the GEV method showing small improvements in mean CRPS compared to the TN method. The regime switching combination method performs best in terms of both mean CRPS and MAE, slightly improving the 
results of the GEV method.  Note that due to the heavier tails, the GEV model generally results in wider prediction intervals compared to the TN model.

 \begin{figure}
 \centering
   \includegraphics[width=0.32\textwidth]{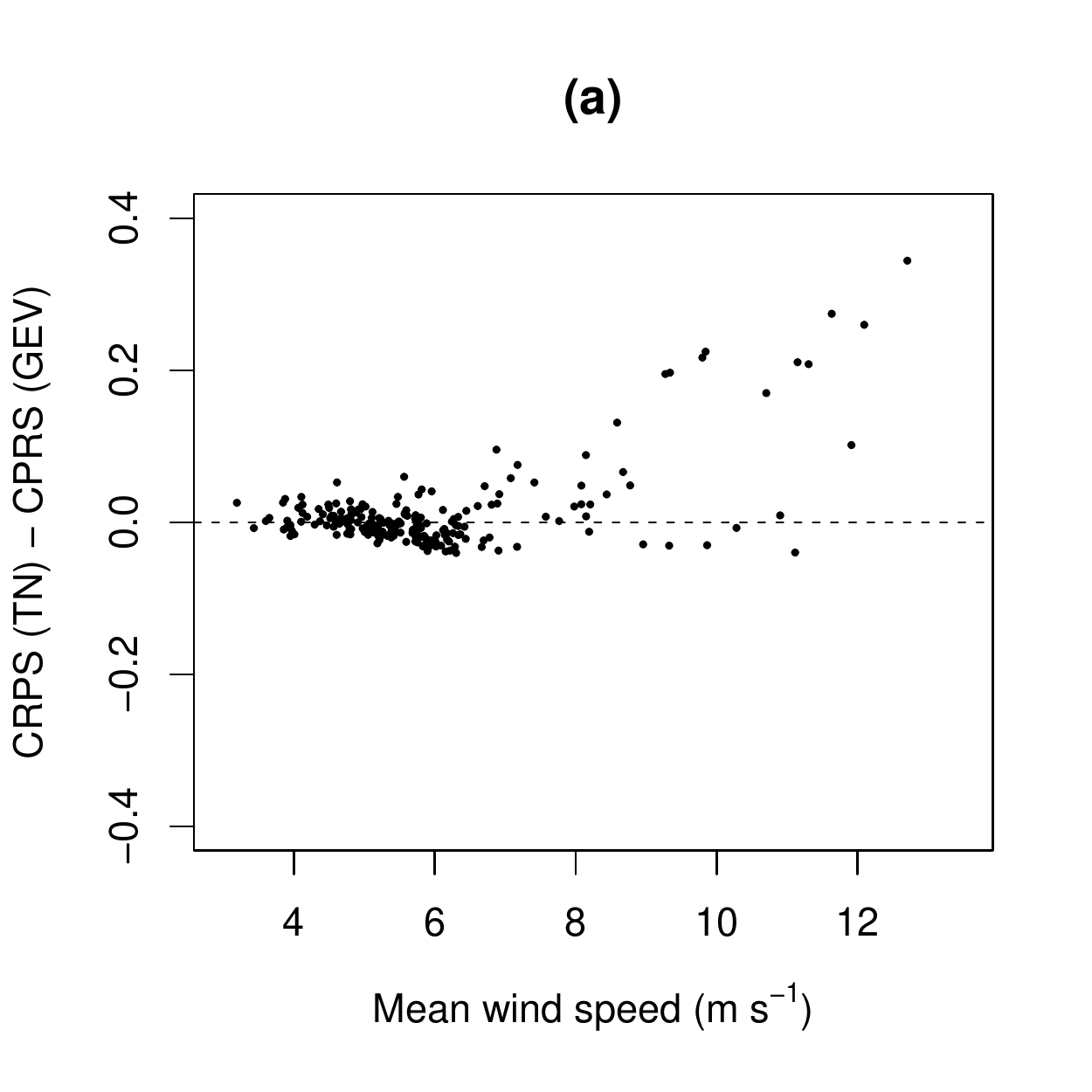}
   \includegraphics[width=0.32\textwidth]{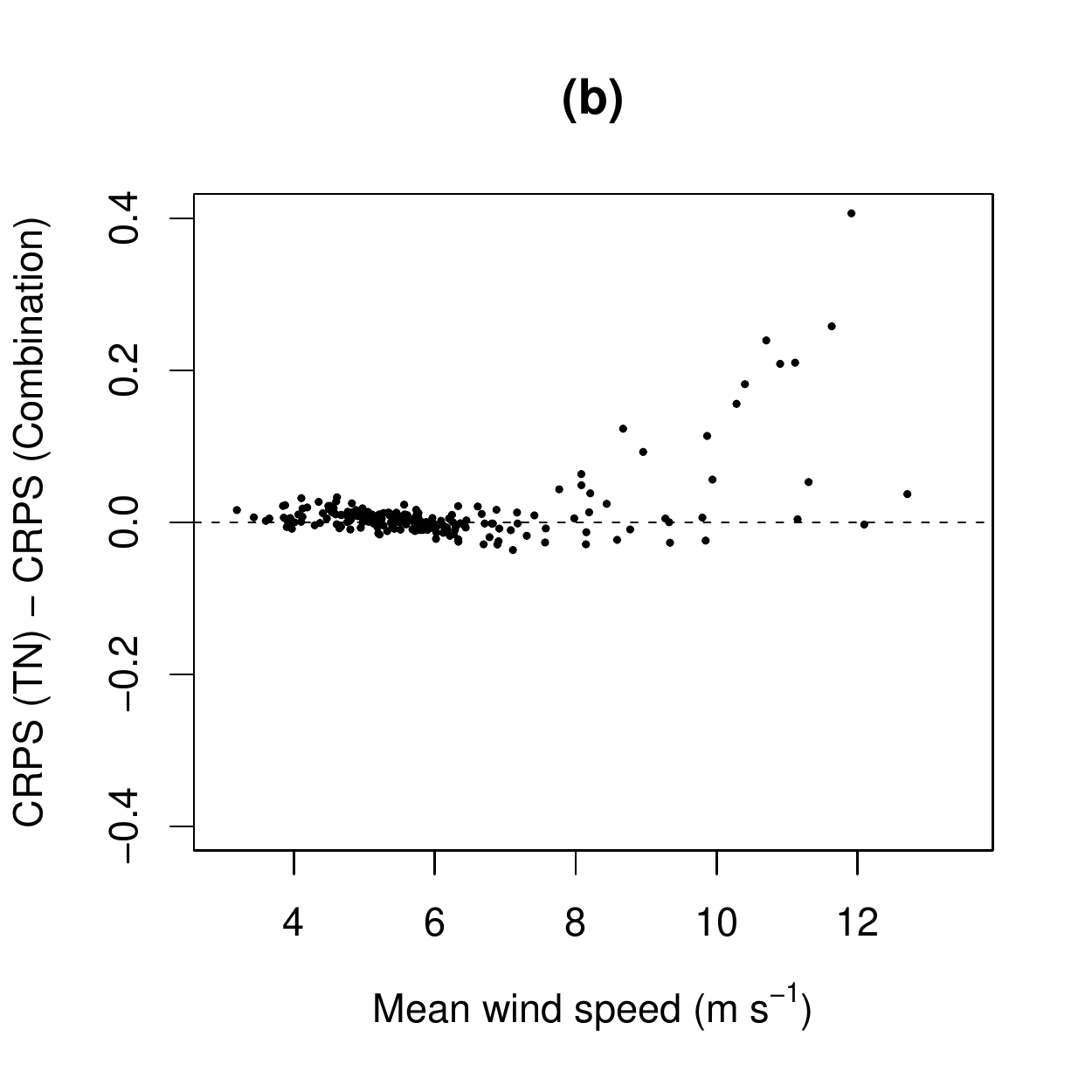}
   \includegraphics[width=0.32\textwidth]{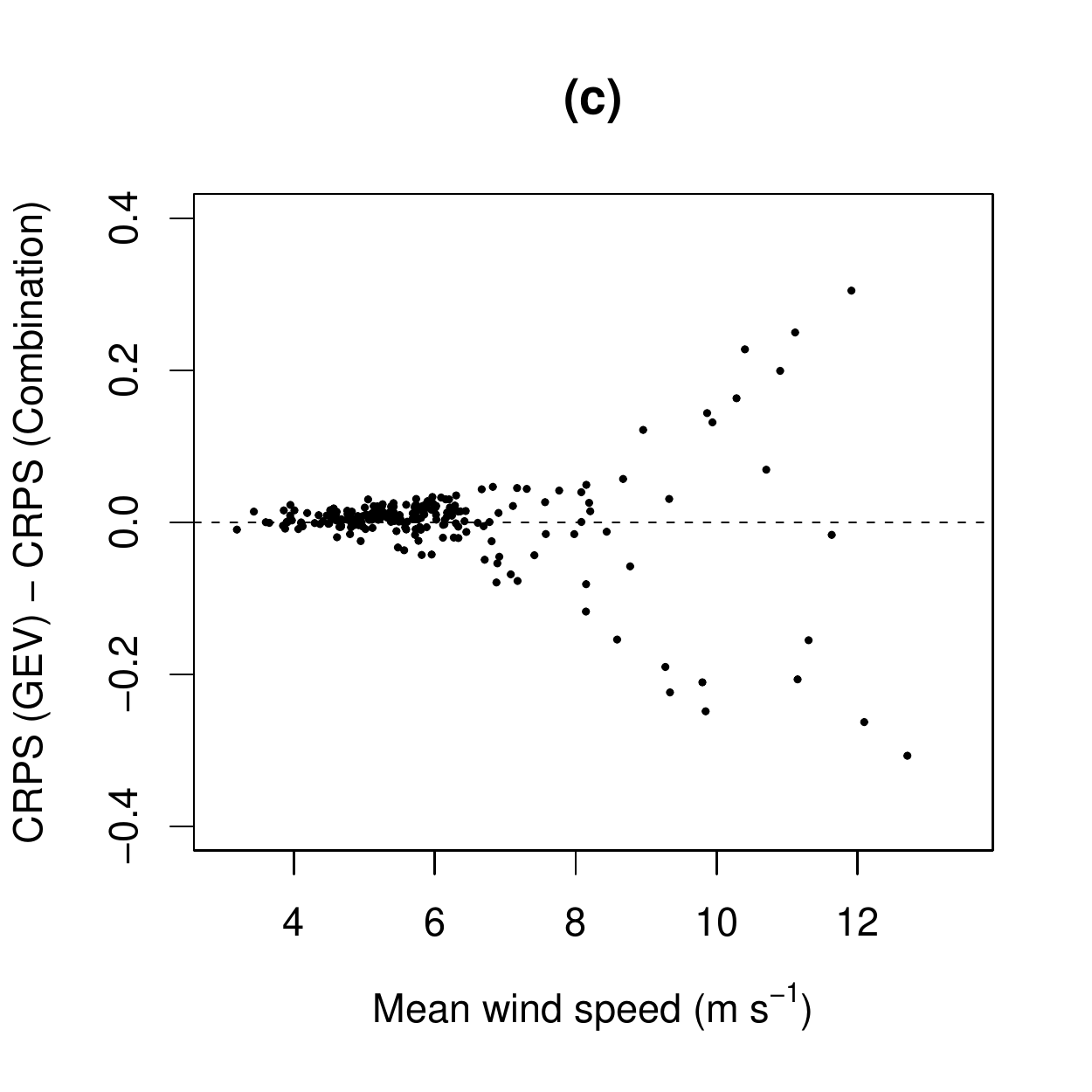} 
   \caption[Station-specific comparison of mean CRPS as a function of the magnitude of average wind speed observations.]{Station-specific comparisons of the continuous ranked probability score (CRPS) for the three postprocessing methods as a function of the average observed daily maximum wind speed at the station.  The plots compare (a) the TN and the GEV model, (b) the TN and the regime switching combination model, and (c) the GEV and the regime switching combination model. The horizontal dashed lines indicate equal predictive performance.}
  \label{fig:crps-diff-by-station}
\end{figure}

  Figure \ref{fig:crps-diff-by-station} compares the station-specific predictive performance of the individual postprocessing models as a function of the site-specific average observed wind speed. Figure~\ref{fig:crps-diff-by-station}(a) and (b) indicate that the overall improvements of the GEV and the regime switching combination model over the TN model are mainly due to improvements at stations with high average observed wind speeds. However,  there appears to be no obvious pattern for stations with high average wind speeds when comparing the GEV model and the regime switching combination model in Figure~\ref{fig:crps-diff-by-station}(c).  The combination model, on the other hand, outperforms the GEV model for most stations with average observed wind speeds below 7 m s$^{-1}$.   

\subsection{One day ahead performance in the upper tail}

  \begin{table}
    \centering
    \caption[Threshold weighted CRPS for one day ahead forecasts.]{Mean threshold weighted continuous ranked probability score (twCRPS) for one day ahead forecasts of daily maximum wind speed at 228 synoptic stations in Germany from 1 May 2010 to 30 April 2011 using an indicator weight function $w_r(z) = \ind\{z \geq r\}$ for different values of $r$.}
    \begin{tabular}{lccc}
      \toprule
      Forecast     & $r = 10$       &  $r = 12$ 	& $r = 15$ \\
      \midrule
      Climatology  & 0.250	     & 0.128 		& 0.045 \\
      Ensemble 	    & 0.211	     & 0.113 		& 0.043 \\
      TN 	    & 0.200	     & 0.111 		& 0.042 \\
      GEV  	    & 0.195	     & 0.107 		& 0.041 \\
      Combination  & \textbf{0.191} & \textbf{0.103} 	& \textbf{0.039}  \\
      \bottomrule
    \end{tabular}
    \label{table:twcrps-24h}
  \end{table}

With a focus on the performance in the upper tail, Table \ref{table:twcrps-24h} shows values of the mean twCRPS for the competing forecasts where we have employed the indicator weight function $w_r(z) = \ind\{z \geq r\}$ for $r = 10, 12$ and $15$ m s$^{-1}$.  The threshold values approximately correspond to the 90th, 95th and 98th percentile of the marginal distribution of the wind speed observations.   The twCRPS is here calculated using numerical integration methods.  All three postprocessing methods improve the ECMWF ensemble predictions, and the GEV approach outperforms the TN method.  The regime switching combination of the two models further improves the performance.  Note that the relative improvement over the TN method for the upper tail is comparatively larger than the improvement under the unweighted CRPS in Table \ref{table:perf-measures-24h}.  Similar rankings hold for any value of $r$ between 10 and 20 m s$^{-1}$.

\begin{figure}
 \centering
   \includegraphics[width=0.5\textwidth]{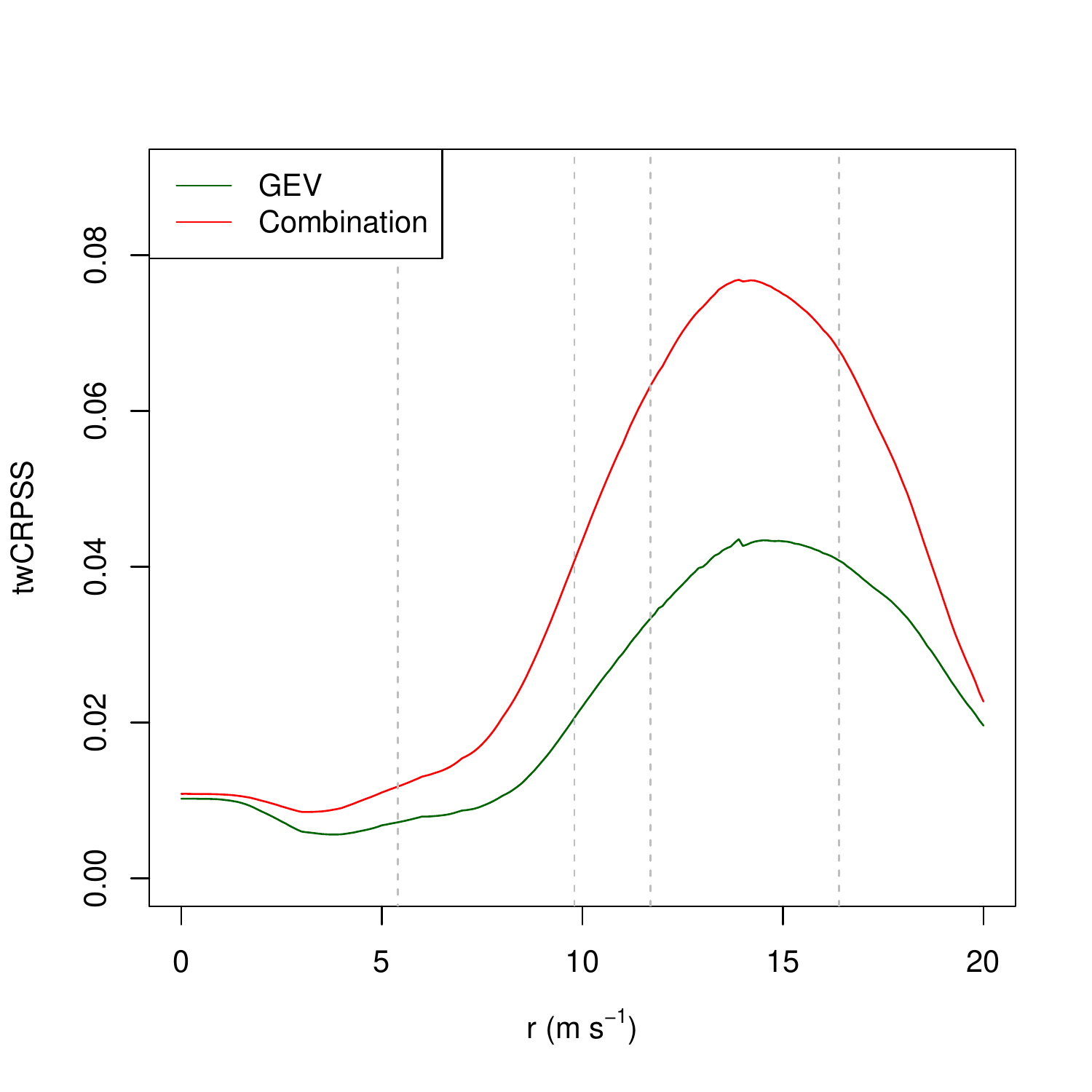}
  \caption[Model comparison for extreme events using the twCRPSS as a function of $r$.]{Threshold-weighted continuous ranked probability skill score of probabilistic one day ahead forecasts of daily maximum wind speed at 228 synoptic stations in Germany from 1 May 2010 to 30 April 2011 as a function of the threshold $r$ in the indicator weight function  $w_r(z) = \ind\{z \geq r\}$, using the forecasts produced by the TN method as reference. The gray dashed vertical lines indicate the 50th, 90th, 95th and 99th percentile of the marginal distribution of the observations.}
  \label{fig:twcrpss}
\end{figure}

  We further consider the threshold-weighted continuous ranked probability skill score (twCRPSS) given by  
\begin{equation}
   \textnormal{twCRPSS}(F,y) = 1 - \frac{\textnormal{twCRPS}(F,y)}{\textnormal{twCRPS}(F_{\textnormal{ref}},y)},
  \end{equation}  
where $F_{\textnormal{ref}}$ denotes the predictive cumulative distribution function of a reference forecast, in our case the TN method. The twCRPSS is positively oriented and can be interpreted as improvement over the reference forecast.  Figure \ref{fig:twcrpss} shows the twCRPSS for the GEV and the regime switching combination method as a function of the threshold $r$ for the indicator weight function, using the TN method as a reference forecast. For all thresholds and both models, the twCRPSS is strictly positive, indicating improved predictive performance compared to the TN model with the regime switching combination method showing greater improvement.  In general, the score values increase for larger threshold values, with the largest differences obtained for threshold values around 14 m s$^{-1}$. 
  
  \subsection{Performance for longer lead times}
 
   \begin{table}
  \centering
  \caption[Summary measures for two and three day ahead forecasts.]{Mean continuous ranked probability score (CRPS), mean absolute error (MAE), average coverage and width of 80\% prediction intervals of daily maximum wind speed forecasts at 228 synoptic stations in Germany from 1 May 2010 to 30 April 2011. The upper half shows results for two day ahead forecasts, the lower half results for three day ahead forecasts.}
  \begin{tabular}{lcccc}
    \toprule
	    & CRPS         & MAE          & Coverage & Width \\
      Forecast & (m s$^{-1}$) & (m s$^{-1}$) & (\%)     & (m s$^{-1}$) \\
    \midrule
    \multicolumn{3}{l}{Two days ahead} \\
    \midrule
      Climatology & 1.55           & 2.14          & 64.4          & 6.6  \\
      Ensemble    & 1.22           & 1.47          & 38.8          & 1.6  \\
      TN          & 1.07           & \textbf{1.43} & \textbf{80.5} & 4.1  \\
      GEV         & \textbf{1.06}  & \textbf{1.43} & 82.5          & 4.7  \\
      Combination & \textbf{1.06}  & \textbf{1.43} & 80.6          & 4.2  \\
    \midrule
    \multicolumn{3}{l}{Three days ahead} \\
    \midrule
      Climatology & 1.55          &  2.14          & 64.4           & 6.6  \\
      Ensemble    & 1.22          &  1.52          & 48.0           & 2.1  \\
      TN          & 1.10          &  \textbf{1.47} & \textbf{80.4}  & 4.3  \\
      GEV         & \textbf{1.09} &  \textbf{1.47} & 82.3           & 4.8  \\
      Combination & \textbf{1.09} &  \textbf{1.47} & 80.7           & 4.3  \\
    \bottomrule
    \end{tabular}
    \label{table:perf-measures-48h72h}
  \end{table}
  
      \begin{table}
    \centering
    \caption[Threshold-weighted CRPS for two and three days ahead forecasts.]{Mean threshold weighted continuous ranked probability score (twCRPS) of daily maximum wind speed forecasts at 228 synoptic stations in Germany from 1 May 2010 to 30 April 2011 using an indicator weight function $w_r(z) = \ind\{z \geq r\}$ for different values of $r$. The upper half shows results for two day ahead forecasts, the lower half shows results for three day ahead forecasts.}
    \begin{tabular}{lccc}
      \toprule
      Forecast     & $r = 10$       &  $r = 12$ 	& $r = 15$ \\
      \midrule
      \multicolumn{3}{l}{Two days ahead} \\
      \midrule
      Climatology   & 0.250	     & 0.128 		& 0.045 \\
      Ensemble 	    & 0.209	     & 0.113 		& 0.043 \\
      TN 	    & 0.202	     & 0.111 		& 0.043 \\
      GEV  	    & 0.198	     & 0.108 		& 0.041 \\
      Combination  & \textbf{0.196} & \textbf{0.106} 	& \textbf{0.040}  \\
      \midrule
      \multicolumn{3}{l}{Three days ahead} \\
      \midrule
      Climatology  & 0.250	     & 0.128 		& 0.045 \\
      Ensemble 	    & 0.209	     & 0.113 		& 0.043 \\
      TN 	    & 0.204	     & 0.112 		& 0.043 \\
      GEV  	    & 0.200	     & 0.109 		& \textbf{0.041} \\
      Combination  & \textbf{0.199} & \textbf{0.107} 	& \textbf{0.041}  \\      
      \bottomrule
    \end{tabular}
    \label{table:twcrps-48h72h}
  \end{table}

 For lead times of two and three days, we obtain similar results as above.  Table \ref{table:perf-measures-48h72h} shows the mean CRPS, MAE and average coverage and width of 80\% prediction intervals for those lead times.  Compared to the one day ahead forecasts, forecasts for longer lead times result in slightly less accurate predictions and wider prediction intervals.  The ECMWF ensemble predictions exhibit wider prediction intervals compared to the one day ahead forecasts resulting in small improvements in calibration.  However, the ensemble predictions are still underdispersive and the three postprocessing methods significantly improve the predictive skill of the ensemble. The differences among the three postprocessing models are less pronounced than for one day ahead forecasts.   Results for the mean twCRPS and lead times of two and three days are shown in Table \ref{table:twcrps-48h72h}.  We obtain the same ranking as for the one day ahead forecasts while the differences in the scores for the three 
models are small.  Note that while the postprocessing methods show a slight decline in accuracy from two to three days ahead, the scores for the upper tail of the ensemble are identical for these lead times.

\section{Discussion}

We propose two extensions to the nonhomogeneous regression ensemble postprocessing approach of \cite{ThoGne2010} employing generalized extreme value predictive distributions for daily maximum wind speed. In a case study over Germany using the exchangeable ECMWF ensemble,  all three nonhomogeneous regression methods significantly improve the skill of the ECMWF ensemble and provide calibrated and accurate probabilistic forecasts.  The best method according to our results is a regime switching method, where a truncated normal model is applied when we expect low winds, while a GEV framework is used when high winds are expected.  

In our GEV approach, we have not accounted for the possibility of the method predicting negative wind speeds.  While this rarely happened in our case study, different results are likely to be obtained for less extreme wind variables.  In an application to quantitative precipitation, \cite{Sch2013} considers the GEV to be left-censored at zero assigning all mass below zero to exactly zero.  This approach seems very appropriate for precipitation, where there is often high probability of zero precipitation.  However, it seems less appropriate for wind variables.  Instead, one might consider a truncation of the GEV distribution similar to the truncated normal distribution in \eqref{eq:TN}.  

The regime-switching combination of the TN and the GEV model offers several starting points for further extensions and potential improvements.   For each forecast case, either the TN or the GEV model is selected based on the median ensemble prediction falling below or exceeding a fixed threshold $\theta$.  Instead of assuming a fixed threshold value, it might be interesting to develop an adaptive method to automatically estimate $\theta$, for example based on station-specific information, or covariate based on other weather variables during the rolling training period.   Alternatively, improvements of the predictive performance for extreme events might be achieved by considering a mixture model using a truncated normal distribution for the bulk of the distribution, and an adaptive generalized Pareto distribution (GPD) for the tail. \cite{BenFri2012} propose such a mixture model for precipitation using lognormal and gamma mixtures for the bulk of the distribution and an adaptive GPD tail which is able to 
significantly improve the predictive performance for extreme quantiles.  Similarly, all three methods could be extended by allowing for local adaption of the parameter estimation.  This might be obtained with a geostatistical approach analogous to \cite{KleEt2011a,KleEt2011b}, where the spatially varying parameters are estimated locally and interpolated to locations without available observations.       

When assessing the predictive performance in the upper tail,  we have focused on the twCRPS with a simple indicator weight functions. Under this weight function, it is not possible to distinguish between forecasts with the same behavior on $[r,\infty)$, but different behavior on $(-\infty,r)$. Following \cite{GneRan2011}, we have also performed our analysis using weight functions of the form $w_r(z) = \Phi(z|\mu_r,\sigma_r^2)$, where $\Phi(\cdot|\mu_r,\sigma_r^2)$ denotes the cumulative distribution function of a normal distribution with mean $\mu_r$ and variance $\sigma_r^2$. For example, we might set $\mu_r = r$ and $\sigma_r^2 = 1$, or $\sigma_r^2 = S^2$, where $S^2$ is the sample variance of the wind speed observations during the out-of-sample training period from 1 February to 30 April 2010. The average values of the twCRPS using these weight functions result in the same ranking and in comparable relative improvements as reported above for the indicator weight function. We have therefore focused on the 
indicator weight function which is computationally less demanding.

\section{Acknowledgments}

We thank Tilmann Gneiting and Michael Scheuerer for helpful discussions and for providing the observation data.  The support of the Volkswagen Foundation through the project ``Mesoscale Weather Extremes--Theory, Spatial Modeling and Prediction (WEX-MOP)'' and Statistics for Innovation $sfi^2$ in Oslo is gratefully acknowledged.

\bibliographystyle{abbrvnat}
\bibliography{references}

\end{document}